\documentclass[twocolumn,           % Format : preprint, twocolumn
               showpacs,            % Pacs : showpacs, noshowpacs
               preprintnumbers,     % Preprint: preprintnumbers,
               aps,                 % Society: ...
               prl,                 % Journal Style : pra, prb, prc, prd, pre,
               a4paper,             % Size : a4paper, ...
               superscriptaddress,  % Affiliation (Title) : groupedaddress,
               nofootinbib,         % Footnote: footinbib, nofootinbib
               tightenlines,        % Remove additional spaces in a line
               floats,floatfix      % Floating Pictures and tables
               ]{revtex4}

\usepackage{graphicx}% Include figure files
\usepackage{dcolumn}% Align table columns on decimal point
\usepackage{bm}% bold math
\usepackage{latexsym}
\usepackage{amsmath,amssymb}        % amssymb includes amsfonts
\usepackage[draft=false]{hyperref}

\usepackage[latin1]{inputenc}
\usepackage{latexsym}
\usepackage{amsmath}
\usepackage{epsfig}
\usepackage{ifpdf}
\begin{document}

\title{Bose gas to Bose--Einstein Condensate by the Phase Transition of the Klein--Gordon equation}

\author{Tonatiuh Matos\footnote{Part of the Instituto Avanzado de Cosmolog\'ia (IAC) collaboration
http://www.iac.edu.mx/}} \email{tmatos@fis.cinvestav.mx}
\affiliation{Departamento de F\'isica, Centro de Investigaci\'on y
de Estudios Avanzados del IPN, A.P. 14-740, 07000 M\'exico D.F.,
M\'exico.}

\author{Elias Castellanos$^*$}
 \email{ecastellanos@fis.cinvestav.mx}
 \affiliation{Departamento de F\'isica, Centro de Investigaci\'on y de Estudios Avanzados
del IPN, A.P. 14-740, 07000 M\'exico D.F., M\'exico.}

\begin{abstract}
We rewrite the complex Klein-Gordon (KG) equation with a mexican-hat
scalar field potential in a thermal bath with one loop contribution
as a new Gross-Pitaevskii-like equation. We interpret it as a
charged and finite temperature generalization of the Gross-Pitaevskii equation. We
find its hydrodynamic version as well and using it, we derive the
corresponding thermodynamics. We obtain a generalized first law for
a charged Bose-Einstein Condensate (BEC). We translate the breaking
of the U(1) local symmetry of the KG field into the new version of
the Gross-Pitaevskii equation and demonstrate that this symmetry breaking
corresponds to a phase transition of the gas into a BEC and show
the conditions for condensation and/or phase transition for which this system naturally becomes superfluid and/or superconductor.
\end{abstract}

\date{\today}
 \maketitle

\section{Introduction}

Since its observation with the help of magnetic traps
\cite{Anderson}, the phenomenon of Bose--Einstein condensation has
spurred an enormous amount of works on the theoretical and
experimental realms associated to this topic. The principal interest
in the study on Bose--Einstein condensation is its interdisciplinary
nature. From the thermodynamic point of view, this phenomenon can be
interpreted as a phase transition, and from the quantum mechanical
point of view as a matter wave coherence arising from overlapping de
Broglie waves of the atoms, in which many of them condense to the
grown state of the system. In quantum field theory, this phenomenon
is related to the spontaneous  breaking of a gauge symmetry
\cite{Bagnato}. Symmetry breaking is one of the most essential
concepts in particle theory and has been extensively used in the
study of the behavior of particle interactions in many theories
\cite{Pinto}. Phase transitions are changes of state, related with changes of symmetries
of the system. The analysis of Symmetry breaking mechanisms have
turn out to be very helpful in the study of phenomena associated to
phase transitions in almost all areas of physics. Bose-Einstein
Condensation is one topic of interest that uses in an extensive way
the Symmetry breaking mechanisms \cite{Bagnato}, its phase
transition associated with the condensation of atoms in the state of
lowest energy and is the consequence of quantum, statistical and
thermodynamical effects.

On the other hand, the results from finite temperature quantum field
theory \cite{Dolan,Weinberg} raise important challenges about their
possible manifestation in condensed matter systems. By investigating
the massive Klein--Gordon equation (KG), in \cite{Matos:2011kn} we
were able to show, that the KG equation can simulate a
condensed matter system. In \cite{Matos:2011kn} it was shown that
the KG equation with a self interacting scalar field (SF)
in a thermal bath reduces to the Gross--Pitaevskii equation in the
no-relativistic limit, provided that the temperature of the thermal
bath is zero. Thus, the KG equation reduces to a
generalized relativistic, Gross--Pitaevskii (GP) equation at finite
temperature. But a question remains open. The KG equation
with a self interacting SF potential defines a symmetry
breaking temperature, at which the system experiments a phase
transition. However, this phase transition does not necessarily
means a condensation of the particles of the system.

In the present work we study the complex KG equation
with a Mexican hat SF potential in a thermal bath. The
idea is very simple, the KG equation up to one loop in perturbations
is able to explain the phase transition of a SF, like the
Higgs field, close to the moment of the phase transition, when the
KG equation breaks the $U(1)$ symmetry of the corresponding
Lagrangian. On the other hand, the Gross-Pitaevskii equation is able to explain
the behavior of a BEC at zero temperature. In this work we wonder if
the KG equation in a thermal bath is able to generalize the Gross-Pitaevskii equation in a region
close to the phase transition of the Bose gas into a BEC, because
the KG equation contains the information of the temperature of the bath in the
SF potential. In base of this idea, we rewrite the complex
KG equation into a Gross-Pitaevskii-like equation an interpret it as a Gross-Pitaevskii one at
finite temperature. In order to see if this equation can explain the
phase transition of a Bose gas into a BEC we write its corresponding
thermodynamics, deriving a corresponding first law for the Bose
gases. The only difference we find here with respect to the
traditional first law of the thermodynamics is a therm where the
quantum mechanical character of the KG equation is present. On the
other hand, a phase transition does no necessarily means gas
condensation. Therefore we obtain the temperature and the conditions
of condensation of this Bose gas. At the end of the work we
qualitatively see under which conditions the Bose gas becomes
superfluid or/and superconductor.

\section{Gauge symmetry breaking}\label{sec:SB}

We start with a model having a local U(1) symmetry given by the
lagrangian %(here $\hat{e}=e/ \hbar c$),
  \begin{eqnarray}
   {\cal L}&=&-\left(\nabla_{\mu}\Phi^*+\mathrm{i}eA_\mu\Phi^*\right)\left(\nabla^{\mu}\Phi-\mathrm{i}eA^\mu\Phi\right)\nonumber\\
   &-&V(|\Phi|) -\frac{1}{4}F^{\mu\nu}F_{\mu\nu}
  \label{eq:L}
  \end{eqnarray}
where $F_{\mu\nu}=\nabla_\mu A_\nu-\nabla_\nu A_\mu$ is the Maxwell
tensor and the SF potential $V$ is the easiest case of a
double-well interacting Mexican-hat potential for a complex SF
$\Phi(\vec{x},t)$, interacting inside a thermal bath in a reservoir
that can have interaction with its surroundings up to one loop of
correction, that goes as
  \begin{eqnarray}
V(\Phi)=-\hat m^2\Phi\Phi^*+\frac{\hat\lambda}{2}(\Phi\Phi^*)^2 &+& \frac{\hat\lambda}{4}k_B^2T^2\Phi\Phi^* \\\nonumber &-& \frac{\pi^2k_B^4}{90\hbar^2c^2}T^4.
\label{eq:V}
 \end{eqnarray}
where $\hat{m}^2=m^2c^2/\hbar^2$ is the scalar particle mass,
$\hat\lambda=\lambda/(\hbar^2c^2)$ is the parameter describing the
interaction, $k_B$ is the Boltzmann's constant, $\hbar$ is the Planck's
constant, $c$ is the speed of light and $T$ is the temperature of
the thermal bath, this result includes both quantum and thermal
contributions.

From here the interpretation of the different quantities of the system is clear, $\Phi$ is the KG version of the Gross-Pitaevskii function $\Psi$ (see bellow), the Maxwell field is an electromagnetic field in the system, $\lambda$ is the self-interaction parameter related with the scattering length of the system and $T$ is the temperature of the thermal bath where the system lie. 

The dynamics of a SF is governed by the KG equation,
it is the equation of motion of a field composed of spinless
particles. In order to confine the BEC experimentally one needs to add an external field controlled by hand in order to cause the condensation, like a laser for example.  In order to do this, we will add an external field $\phi$ that will interact with the SF to first
order, such that the KG equation will be given
by
 \begin{equation}
 \Box_E^{2}\Phi-\frac{dV}{d\Phi^*}-2\hat m^2\phi{\Phi}=0,
 \label{eq:KG}
 \end{equation}
where for a charged field the D'Alambertian operator is given by,
 \begin{equation}
  \Box_E
 ^{2}\equiv\left(\nabla_{\mu}-\mathrm{i}eA_\mu\right)\left(\nabla^{\mu}-\mathrm{i}eA^\mu\right)
  \label{eq:Box}
 \end{equation}
where $A_\mu$ is the electromagnetic four potential. Observe that we can rewrite the D'Alambertian as
\begin{equation}
\Box_E^2=(\boldsymbol{\nabla}-2\mathrm{i}
e\boldsymbol{A})\cdot\boldsymbol{\nabla}-(\frac{1}{c}\frac{\partial}{\partial
t}-2\mathrm{i}e\varphi)\frac{1}{c}\frac{\partial}{\partial
t}-\mathrm{i}e\nabla_\mu A^\mu-e^2A_\mu A^\mu
\label{eq:Box1}
\end{equation}
where we have written the electromagnetic four potential as
$A_\mu=(\boldsymbol{A},\varphi)$.
In what follows we will use the Lorentz gauge $\nabla_\mu A^\mu=0$.
It is convenient to consider the total (effective) potential $V$ adding the
external one and the term $e^2A_\mu A^\mu=e^2 A^2 $ to the potential,
such that
\begin{eqnarray}
V_T(\Phi)=&-&\frac{m^2c^2}{\hbar^2}\Phi\Phi^*+\frac{\lambda}{4\hbar^2c^2}k_B^2T^2\Phi\Phi^*-e^2A^2\Phi\Phi^*\nonumber\\&+&\frac{\lambda}{2\hbar^2c^2}(\Phi\Phi^*)^2 -\frac{\pi^2k_B^4}{90\hbar^2c^2}T^4+\frac{2m^2c^2}{\hbar^2}\phi{\Phi\Phi^*}.\nonumber\\
\label{eq:VT}
\end{eqnarray}
where we see that the parameter $e$ is the coupling constant between the electromagnetic and the SFs. In terms of the effective potential $V_T$ the KG equation now can be written as
 \begin{equation}
 \Box^2\Phi-ieA^\mu\nabla_\mu\Phi-\frac{dV_T}{d\Phi^*}=0,
 \label{eq:KGT}
 \end{equation}
 where now $\Box^2=\nabla^\mu\nabla_\mu$. The Maxwell equations also read
 \begin{eqnarray}
 \nabla^\mu F_{\mu\nu}&=&-j_\nu\nonumber\\
 &=&ie(\Phi^*\nabla_\nu\Phi-\Phi\nabla_\nu\Phi^*)+2e^2\Phi\Phi^*A_\nu.
 \label{eq:Mawxell}
 \end{eqnarray}

 We can define an effective mass by
 \begin{equation}
 m_{eff}c^2=\sqrt{m^2c^4+\hat{e}^2A^2}.
 \label{eq:KGT1}
 \end{equation}
 where $\hat e=e\hbar c$.
 %Here the term with the electromagnetic potential is the Proca mass.
For the $V_T$ potential \eqref{eq:VT}, the critical temperature
$T^{SB}_{c}$  where the minimum of the potential $\Phi=0$ becomes a
maximum and at which the symmetry is broken is,
 \begin{equation}
  k_B T^{SB}_{c}=\frac{2c^2}{\sqrt{\lambda}}\sqrt{m_{eff}^2-2 m^2\phi}.
 \label{eq:Tc}
 \end{equation}
Potential \eqref{eq:VT} has a minimum in $\Phi=0$ when the
temperature $T>T^{SB}_{c}$. If $T<T^{SB}_{c}$, the point $\Phi=0$
becomes a maximum and potential  \eqref{eq:VT} has two minima in
 \begin{eqnarray}
  R_{min}&=&\pm\sqrt{\frac{1}{\lambda}(m^2c^4+\hat{e}^2A^2 -\frac{\lambda}{4}k_B^2 T^2-2 m^2c^4\phi)}\nonumber\\
  &=&\pm\frac{k_B}{2}\sqrt{(T^{SB}_{c})^2- T^2}
 \label{eq:min}
 \end{eqnarray}
being $\Phi=Re^{i\Theta}$. In the maximum $\Phi=0$ the second derivative of the potential $V_T$
with respect to the SF reads
 \begin{eqnarray}
  V_{T,\Phi\Phi}&=&-\,\left(\hat m^2+e^2A^2-\frac{\hat\lambda}{4} k_B^2T^2-2\hat m^2\phi\right)\nonumber\\
  &=&-\frac{\hat\lambda}{4}k_B^2\left((T^{SB}_{c})^2-T^2\right)\nonumber\\
  &=&-\,\frac{(T^{SB}_{c})^2-T^2}{(T^{SB}_{c})^2}\left(\hat m_{eff}^2-2\hat m^2\phi\right)c^4.
 \label{eq:min1}
 \end{eqnarray}

In what follows we will re-write the KG equations in order to interpret the KG equation as a Gross-Pitaevskii one.  

\section{The generalized Gross-Pitaevskii equation}\label{sec:GP}

Now for the SF we perform the transformation
 \begin{equation}
\Phi=\Psi\,\mathrm{e}^{-\mathrm{i}\hat mct}\label{eq:Phi}
  \nonumber,
 \end{equation}

In terms of the complex function $\Psi$, the KG equation
(\ref{eq:KG}) now reads,
\begin{eqnarray}
 \mathrm{i}\hbar c\dot{\Psi}+\frac{\hbar^2}{2m}\Box_E^2{\Psi}&-&
\frac{\lambda}{2mc^2}|\Psi|^2\Psi-[mc^2(\phi-1)+ec\varphi]\Psi\nonumber\\
 &-&\frac{\lambda k_B^2T^2}{8mc^2}\Psi=0,
\label{eq:GP}
\end{eqnarray}
where we have kept just the equation for the $\Psi$ part, the
complex conjugate can be described in the same way. The notation
used is  $\dot{}=1/c\,\partial/\partial t$  
(\ref{eq:GP}) is the KG equation \eqref{eq:KG} or \eqref{eq:KGT}
rewritten in terms of the function $\Psi$ and temperature $T$. This
equation is an exact equation defining the field
$\Psi(\mathbf{x},t)$, where $\phi$ defines the external potential
acting on the system and the term with $\lambda$ represents the
self-interaction of the SF within the system. We will consider equation
(\ref{eq:GP}) as a generalization of the Gross-Pitaevskii equation
for finite temperatures and relativistic particles. This is because
 when $T=0$ and in the non-relativistic limit,
$\Box^2\rightarrow\nabla^2$, eq. (\ref{eq:GP}) becomes the
Gross-Pitaevskii equation for Bose-Einstein Condensates (BEC),
provided that $\lambda=8\pi\hbar^2c^2 a\,\kappa^2$, being $a$ the s-wave
scattering length \cite{Pethick}. The static limit of equation
(\ref{eq:GP}) is known as the Ginzburg-Landau equation. 
Observe that the temperature $T$ and the external field $\phi$
must be manipulated from outside.

At this point it is noteworthy to mention the order of magnitude of the quantity \eqref{eq:Tc}, the associated critical temperature at which the symmetry of the system is broken. Assuming that the external potential $\phi$ and the additional field $A^{2}$ are time independent, i.e, that the system is in static thermal equilibrium, we can write $k_BT^{SB}_{c}\sim\frac{2mc^{2}}{\sqrt{\lambda}}$ in the center of the system.  If we set $\lambda=8\pi\hbar^{2}c^{2}a\kappa^{2}$ into \eqref{eq:Tc}, thus $T^{SB}_{c}\sim8.21\frac{Joules^{-1}\text{m}^{-3/2}}{\kappa}\left(\frac{m}{\text{gr}}\right)\sqrt{\frac{\text{cm}}{a}}\times 10^{62}\,K$.
%, which implies that the corrections in the condensation temperature caused by the thermal bath in the non--relativistic limit cannot be negligible, provided that the quantity $\frac{m}{\kappa \sqrt{a}}\sim 10^{-73}$. In this case, the critical temperature for SB is of the order of nano--kelvins, suggesting that the \emph{scale} $\kappa$ must be very large under typical conditions, $\kappa \sim \frac{m}{\sqrt{a}}10^{73}$. 
For instance, in the case of $^{87}$Rb, with a mass $m\sim 1.452\times10^{-22}\,$gr and a scattering length $a\sim5.2\times10^{-7}$m, the critical temperature of symmetry breakdown of this system is $T^{SB}_{c}\sim\frac{1.65}{\kappa}\times10^{44}\,K$, which depending on the value of $\kappa$, it could be very big. However, for example, if we set
$\kappa\sim5\times10^{50}$ $Joules^{-1}\,$m$^{-3/2}$ (see also \cite{Castellanos:2012nr}), we obtain that $T^{SB}_{c}\sim3\times10^{-7}\,K$. Nevertheless, the density $n$ does not depend on the value of $\kappa$, because $n=\kappa^2\,k_B\,(T_c^{SB})^2/4(1-({T}/{T_c^{SB}})^2)\sim\kappa^2\,k_B\,(T_c^{SB})^2/4=m^2c^2/(8\pi\hbar^2\,a)$, and this quantity for the $^{87}$Rb is $\sim10^{36}/$cm$^3$. Multiplying this density times the mass, we obtain the corresponding mass density, which is $\sim1.9\times10^{14}$gr$/$cm$^3$. This mass density corresponds to the one of a nuclear atom or a neutron star. Furthermore, for this values the self-interaction parameter $\lambda$ is very big $\lambda\sim3.2\times10^{43}$.  Let us suppose for a moment that neutrons can build cooper pairs in a neutron star. In that case, the mass of two neutrons is $m=2\times1.6\times10^{-24}$gr, the scattering length of the neutrons is $a\sim10^{-11}$cm. Thus, the critical temperature of symmetry breakdown of this system is $T^{SB}_{c}\sim\frac{8.31}{\kappa}10^{44}\,K$. If we set the typical temperature of collapse for  a  neutron star $T\sim10^{12}$K as the temperature of the symmetry breakdown, we obtain that $\kappa\sim10^{32}$ $Joules^{-1}\,$m$^{-3/2}$. In this case $\lambda\sim25$, the particle density $n\sim3.2\times10^{37}/$cm$^3$ and the mass density $\sim10^{14}$gr$/$cm$^3$, which is again of the order of magnitude of neutron stars (see \cite{dany}). Other interesting example is BECs as Dark Matter \cite{juan}. In this case the mass of the SF could be of the order of the axion mass $m\sim0.1$eV$=1.78\times10^{-34}$gr. Here there are two cases, if the SF is an axion \cite{sikivie}, the self-interaction parameter is $\lambda\sim10^{15}$ \cite{barranco}, for this value of the mass and the self-interacting parameter, the critical temperature is $T^{SB}_c\sim7\times10^{-5}$K$=6\times10^{-9}$eV. The mass density is as expected $\rho\sim10^{-45}$gr/cm$^3$. On the other hand, if the SF are Scalar Field Dark Matter particles \cite{victor}, the mass is again  $m\sim0.1$eV$=1.78\times10^{-34}$gr, but the self-interaction parameter is $\lambda\sim10^{-6}$. For that the critical temperature is $T^{SB}_c\sim2.3\times10^{6}$K$=200$eV for a mass density of the order of the critical density of the universe $\rho\sim10^{-29}$gr/cm$^3$. 

\section{The Hydrodynamical version}\label{sec:hydro}

In what follows we transform the generalized Gross-Pitaevskii
equation (\ref{eq:GP}) into its analogous hydrodynamical version,
\cite{Chiueh,Bohm}, for this purpose the ensemble wave function
$\Psi$ will be represented in terms of a modulus $n$ and a phase $S$
as,
 \begin{equation}
 \kappa \Psi=\sqrt{ n}\,\mathrm{e}^{\mathrm{i}S}.
 \label{eq:psi}
 \end{equation}
where the phase $S(\boldsymbol{x},t)$ is taken as a real function.
As usual this phase will define the velocity. Here we will interpret
$n(\boldsymbol{x},t)$ as the number
density of particles in the symmetry broken state, such that
$\kappa^2|\Psi|^2=\kappa^2\Psi\Psi^*=n$, where $\kappa$ is the scale
of the system, which is to be determined by an experiment, being both, $S$ and $
n$, functions of time and position.
So, from this interpretation we have that when the KG equation
oscillates around the $\Phi=0$ minimum, the number of particles in
the ground state is zero, $n=0=\rho$. Below the critical temperature
$T^{SB}_{c}$, close to the second minimum,
$R_{min}^2=k_B^2((T^{SB}_{c})^2-T^2)/4$, the density will
oscillate around $n=\kappa^2 k_B^2((T^{SB}_{c})^2-T^2)/4$ as can be
seen by equation (\ref{eq:psi}). In order to see this more clear, we
perform the Madelung transformation \eqref{eq:psi} in the
generalized Gross-Pitaevskii equation \eqref{eq:GP}.

 From (\ref{eq:GP}) and (\ref{eq:psi}), separating real and imaginary parts we obtain
 \begin{subequations}
 \begin{eqnarray}
 c\,\dot{ n}+\frac{\hbar}{m}n\left[ \Box^2 S-e\left(\nabla\cdot\boldsymbol{A}-\dot\varphi\right)\right]&+&
\nonumber\\
 \frac{\hbar}{m}\left((\boldsymbol{\nabla}S-e\boldsymbol{A})\cdot\boldsymbol{\nabla} n
  -(\dot S-e\varphi)\dot{ n}\right)&=&0\label{eq:ntotal},
  \end{eqnarray}
   \begin{eqnarray}
 \frac{\hbar c}{m}\,(\dot S-e\varphi) +\frac{\lambda}{2m^2c^2\kappa^2} n+c^2(\phi-1) +\frac{\lambda}{8m^2c^2}k_B^2T^2&+&\nonumber\\
  \frac{\hbar^2}{m^2}\left(\frac{\Box^2\sqrt{ n}}{\sqrt{ n}}\right)+\frac{\hbar^2}{2m^2}\left((\boldsymbol{\nabla}S-e\boldsymbol{A})^2-(\dot S-e\varphi)^2\right)&=&0.\nonumber\\
   \label{eq:Stotal}
 \end{eqnarray}\label{eq:hidro}
  \end{subequations}
Taking the gradient of (\ref{eq:Stotal}) and using the definitions
of the fluxes
\begin{subequations}
\begin{eqnarray}
\boldsymbol{ j}&=&\frac{2en}{\kappa^2}(\boldsymbol{\nabla} S-e\boldsymbol{A})\label{eq:fluxesjvec}\\
j&=&\frac{2en}{\kappa^2}(\dot S-e\varphi)\label{eq:fluxesj}\\
j_\mu&=&(\boldsymbol{j},j-\frac{2en}{\kappa^2}\hat m)
\label{eq:fluxesjmu}
\end{eqnarray}
\label{eq:fluxes}
\end{subequations}
and the velocity
 \begin{equation}
  \boldsymbol{v}\equiv\frac{\hbar}{m}\left(\boldsymbol{\nabla}S-e\boldsymbol{A}\right)
 \label{eq:vel}
 \end{equation}
equations \eqref{eq:hidro} can be rewritten as,
\begin{subequations}
 \begin{eqnarray}
  \dot{ n}+\boldsymbol{\nabla}\cdot( n\boldsymbol{v})
  -\frac{\hbar\kappa^2}{2me c}\dot j&=&0,\label{eq:cont}\\
  \dot{\boldsymbol{v}}+(\boldsymbol{v}\cdot\boldsymbol{\nabla})\boldsymbol{v}-\frac{\hbar}{m}e\left(c\boldsymbol{E}+\boldsymbol{v}\times\boldsymbol{B}\right)&=&\nonumber\\
  -c^2\boldsymbol{\nabla}\phi
  -\frac{\lambda}{m^2c^2\kappa^2}\boldsymbol{\nabla} n
  -\frac{\hbar^2}{m^2}\boldsymbol{\nabla}\left(\frac{\nabla^2\sqrt{ n}}{\sqrt{ n}}\right)&+&\nonumber\\
 \frac{\hbar^2}{2m^2}\boldsymbol{\nabla}(\dot S-e\varphi)^2 +\frac{\hbar^2}{m^2}\boldsymbol{\nabla}\left(\frac{\partial^2_t\sqrt{ n}}{\sqrt{ n}}\right)
  &-&\frac{\lambda k^2_B}{4m^2}T\boldsymbol{\nabla}T\nonumber\\
  \label{eq:2}
 \end{eqnarray}\label{eq:hydro}
\end{subequations}
where $\boldsymbol{E}=-1/c\partial\boldsymbol{A}/\partial
t+\nabla\cdot\varphi$ and
$\boldsymbol{B}=\nabla\times\boldsymbol{A}$ respectively are the
electric and the magnetic field vectors. Notice that in (\ref{eq:2})
$\hbar$ enters on the right-hand side through the term containing
the gradient of $ n$. This term is usually called the 'quantum
pressure' and is a direct consequence of the Heisenberg uncertainty
principle, it reveals the importance of quantum effects in
interacting gases. Multiplying by $ n$, (\ref{eq:2}) can be written
as:
 \begin{eqnarray}
  n\dot{\boldsymbol{v}}+n(\boldsymbol{v}\cdot\boldsymbol{\nabla})\boldsymbol{v}=n\boldsymbol{F}_E+n\boldsymbol{F}_\phi -\boldsymbol{\nabla}p
  +n\boldsymbol{F}_Q+\boldsymbol{\nabla}\sigma,
 \label{eq:navier}
 \end{eqnarray}
where $\boldsymbol{F}_E=\frac{e}{m}
\left(c\boldsymbol{E}+\boldsymbol{v}\times\boldsymbol{B}\right)$ is
the electromagnetic force,
$\boldsymbol{F}_\phi=-\boldsymbol{\nabla}\phi$ is the force
associated to the external potential $\phi$, $p$ can be seen as the
pressure of the SF gas that satisfies the equation of state
$p=wn^2$, being $\omega=\lambda/(2m^2c^2\kappa^2)$ an interaction
parameter. $\boldsymbol{\nabla}p$ are forces due to the gradients of
pressure, $\boldsymbol{F}_Q=-\boldsymbol\nabla U_Q$ is the quantum
force associated to the quantum potential, \cite{gro,Pethick},
\begin{equation}
 U_Q=\frac{\hbar^2}{m^2}\left(\frac{\nabla^2\sqrt{ n}}{\sqrt{ n}}\right),
 \label{eq:UQ}
 \end{equation}
and $\boldsymbol{\nabla}\sigma$ is defined as
\begin{eqnarray}
 \boldsymbol{\nabla}\sigma= \frac{\hbar^2}{2m^2}n\boldsymbol{\nabla}(\dot S-e\varphi)^2
 &-&\frac{1}{4}\frac{\lambda}{m^2}k_B^2 nT\boldsymbol{\nabla}T\nonumber\\
 -\zeta\boldsymbol{\nabla}(\ln n{\dot)}
 &+&\frac{\hbar^2n}{2m^2}\boldsymbol{\nabla}\left(\frac{\ddot{n}}{n}\right),
 \label{eq:sigma}
 \end{eqnarray}
where the coefficient $\zeta$ is given by
\begin{equation}
 \zeta=\frac{\hbar^2}{2m^2}\left[-\boldsymbol\nabla\cdot( n\boldsymbol{v})+\frac{\hbar\kappa^2}{2mec}\dot{j}\right]\nonumber.
 \label{eq:zeta}
 \end{equation}
Observe that using \eqref{eq:cont} the term $\boldsymbol\nabla(\ln n{\dot)}$ can be written as
\begin{eqnarray}
 \boldsymbol{\nabla}(\ln n{\dot)}=-\boldsymbol{\nabla}(\boldsymbol{\nabla}\cdot\boldsymbol{v})
 -\boldsymbol{\nabla}[\boldsymbol\nabla(\ln n)\cdot\boldsymbol{v}]
 +\frac{\hbar\kappa^2}{2mec}\boldsymbol{\nabla}\left(\frac{1}{ n}\dot j\right).\nonumber
 \label{eq:vrho1}
 \end{eqnarray}
System \eqref{eq:hydro} is the hydrodynamical representation to
equation (\ref{eq:GP}) and up to one constant, it is completely equivalent to (\ref{eq:GP}).

\section{The Newtonian (non-relativistic) limit}\label{sec:Newton}

  Neglecting second order time derivatives and products of time derivatives we can simplify system (\ref{eq:hydro}). In
this limit we arrive to the non-relativistic system of equations
\eqref{eq:hydro},
\begin{subequations}
 \begin{eqnarray}
  \dot{ n}+\boldsymbol{\nabla}\cdot( n\boldsymbol{v})&=&0,\label{eq:contNR}\\
  n\dot{\boldsymbol{v}}+n(\boldsymbol{v}\cdot\boldsymbol{\nabla})\boldsymbol{v}&=&n\boldsymbol{F}_E+n\boldsymbol{F}_\phi
  -\boldsymbol{\nabla}p+n\boldsymbol{F}_Q+\boldsymbol{\nabla}\sigma.\nonumber\\
   \label{eq:navierNR}
 \end{eqnarray}\label{eq:hydroNR}
 \end{subequations}
Equation (\ref{eq:contNR}) is the continuity equation, and
(\ref{eq:navierNR}) is the equation for the momentum. Observe that
this last one contains forces due to the external potential, to the
gradient of the pressure, viscous forces due to the interactions of
the condensate and forces due to the quantum nature of the
equations. Quantity $\boldsymbol{\nabla}(\ln n{\dot)}$ plays a very
important roll, in this limit it reads
\begin{equation}
 \boldsymbol{\nabla}(\ln n{\dot)}=-\boldsymbol{\nabla}(\boldsymbol{\nabla}\cdot\boldsymbol{v})
 -\boldsymbol{\nabla}[\boldsymbol{\nabla}(\ln n)\cdot\boldsymbol{v}]\nonumber.
 \label{eq:vrho1NR}
 \end{equation}
 Thus
\begin{eqnarray}
 \boldsymbol{\nabla}\sigma=-\frac{1}{4}\frac{\lambda}{m}k_B^2 nT\boldsymbol{\nabla}T
 -\zeta\left[\boldsymbol{\nabla}(\boldsymbol{\nabla}\cdot\boldsymbol{v})
 +\boldsymbol{\nabla}[\boldsymbol{\nabla}(\ln n)\cdot\boldsymbol{v}]\right],
\label{eq:sigma2}
 \end{eqnarray}
where now we have
\begin{equation}
 \zeta=-\frac{\hbar^2}{2m^2}\boldsymbol{\nabla}\cdot( n\boldsymbol{v})\nonumber,
 \end{equation}
We interpret the function $\boldsymbol{\nabla}\sigma$ as the
viscosity of the system, it contains terms which are gradients of
the temperature and of the divergence of the velocity and density
(dissipative contributions). The measurement of the temperature
dependence in this thermodynamical quantity at the phase transitions
might reveal important information about the behavior of the gas due
to particle interaction.

\section{The Thermodynamics}\label{sec:thermodynamic}

In what follows we will derive the thermodynamical equations for the non-relativistic limit from
the hydrodynamical representation. We can derive a conservation
equation for a function $\alpha$, starting with the relationship
 \begin{equation}
  ( n\alpha)\dot{}= n\dot{\alpha}+\alpha\dot{n}
 \label{eq:phi1}
 \end{equation}
where $\alpha$ can take the values of $\phi$, $U_Q$, etc., all of them
fulfil equation (\ref{eq:phi1}). Using the continuity equation
(\ref{eq:contNR}) in \eqref{eq:phi1} we obtain,
  \begin{equation}
   ( n\alpha)\dot{}+\boldsymbol{\nabla}\cdot(n\boldsymbol{v}\alpha)=-n\boldsymbol{v}\cdot\boldsymbol{F}_{\alpha}
   +n\dot{\alpha}\nonumber.
   \label{eq:contphi}
  \end{equation}
Nevertheless, this procedure is not possible for $\sigma$ because in
general we do not know it explicitly, only in some cases it might be
possible to integrate it.

 Observe how the quantum potential $U_Q$ also fulfills the following relationship
  \begin{equation}
   n \dot U_Q+\boldsymbol{\nabla}\cdot( n\boldsymbol{v}_\rho)=0,
   \label{eq:contUQ21}
  \end{equation}
which follows by direct calculation, and where we have defined the
velocity density $\boldsymbol{v}_\rho$ by
  \begin{equation}
  \boldsymbol{v}_\rho=\frac{\hbar^2}{4m^2}\left(\boldsymbol{\nabla}\ln n\right)\dot{}\nonumber,
  \label{eq:vrho}
  \end{equation}
which can be interpreted as a velocity flux due to the potential
$U_Q$.  Using the continuity equation \eqref{eq:contNR}, equation
(\ref{eq:contUQ21}) can be rewritten as
  \begin{equation}
   (n U_Q)\dot{}+\boldsymbol{\nabla}\cdot (n U_Q\boldsymbol{v}+\boldsymbol{J}_\rho)+ n\boldsymbol{v}\cdot\boldsymbol{F}_Q=0
   \label{eq:contUQ2}
  \end{equation}
where we have  defined the quantum density flux
  \begin{equation}
  \boldsymbol{J}_\rho= n\boldsymbol{v}_\rho\nonumber.
   \label{eq:Jrho}
  \end{equation}
Equation \eqref{eq:contUQ2} is another expression for the continuity
equation of the quantum potential $U_Q$.

 As we know, for the non-relativistic limit the total energy density of the system $\epsilon$ is the sum of the kinetic,
potential and internal energies \cite{oli}, in this case we have an
extra term $U_Q$ due to the quantum potential,
 \begin{equation}
  \epsilon=\frac{1}{2} nv^2+ n\phi+ nu+ nU_Q+\psi_E
 \label{eq:energiae}
 \end{equation}
being $u$ the inner energy of the system and
 \begin{equation}
  \psi_E=\frac{e}{m}(\varphi-\boldsymbol{v}\cdot\boldsymbol{A})
 \label{eq:psiE}
 \end{equation}
the electromagnetic energy potential, defined in terms of the vector
potential $\boldsymbol{A}$ and the electric potential $\varphi$.
Observe that $\psi_E$ fulfills the continuity equation
 \begin{equation}
  (n\psi_E\dot)+\boldsymbol{\nabla}\cdot(n\boldsymbol{v}\psi_E+\boldsymbol{j}_B)=n\dot\psi_E-n\boldsymbol{v}\cdot \boldsymbol{F}_E
 \label{eq:psiE1}
 \end{equation}
being $\boldsymbol{j}_B$ given by the continuity equation of the
vector potential $\boldsymbol{A}$
\begin{equation}
 \frac{\partial \boldsymbol{A}}{\partial t}+(\boldsymbol{v}\cdot\boldsymbol{\nabla}) \boldsymbol{A}=-(\boldsymbol{A}\cdot\boldsymbol{\nabla})\boldsymbol{v}+\frac{m}{e}\boldsymbol{j}_B,
   \label{eq:contA}
  \end{equation}

 Then from (\ref{eq:energiae}) we have that $u$ will satisfy the equation
  \begin{equation}
   ( nu)\dot{}+\boldsymbol{\nabla}\cdot\boldsymbol{J}_u-\boldsymbol{\nabla}\cdot\boldsymbol{J}_{\rho}+ n\dot{\phi}=
   -p\boldsymbol{\nabla}\cdot\boldsymbol{v},
    \label{eq:conte}
  \end{equation}
being $\boldsymbol{J}_u$ the energy current,  given by a energy flux
and a heat flux, $\boldsymbol{J}_q$,
  \begin{equation}
   \boldsymbol{J}_u=n u\boldsymbol{v}+\boldsymbol{J}_q+\boldsymbol{J}_B-p\boldsymbol{v},\nonumber
   \label{eq:Je}
  \end{equation}
where
$\boldsymbol{\nabla}\cdot\boldsymbol{J}_q=\boldsymbol{v}\cdot(\boldsymbol{\nabla}\sigma)$,
and
$\boldsymbol{\nabla}\cdot\boldsymbol{J}_B=\boldsymbol{v}\cdot(n\boldsymbol{j}_B)$,
expressions that as we can see are related in a direct way to the
velocity and gradients of temperature in the condensate, and are the
ones that show in an explicit way the temperature dependence of the
thermodynamical equations. With these definitions at hand we have,
  \begin{equation}
   \left(n u\right)\dot{}+\boldsymbol{\nabla}\cdot(n\boldsymbol{v}u+\boldsymbol{J}_q+\boldsymbol{J}_B-p\boldsymbol{v}
   -\boldsymbol{J}_\rho)+ n\dot{\phi}=-p\boldsymbol{\nabla}\cdot\boldsymbol{v}.
   \label{eq:contu}
  \end{equation}

In order to find the thermodynamical quantities of the system in
equilibrium (taking $p$ as constant on a given volume), we restrict
the system to the regime where the auto-interacting potential is
constant in time, with this conditions at hand for (\ref{eq:contu})
we have:
\begin{equation}
   \left(n u\right)\dot{}+\boldsymbol{\nabla}\cdot(n\boldsymbol{v}u+\boldsymbol{J}_q+\boldsymbol{J}_B-p\boldsymbol{v}-\boldsymbol{J}_\rho)=
   -p\boldsymbol{\nabla}\cdot\boldsymbol{v}
   \label{interna}
  \end{equation}
From (\ref{interna}) we can have a straightforward interpretation of
the terms involved in the phase transition. As always the first term
will represent the change in the internal energy of the system,
$-p\boldsymbol{\nabla}\cdot\boldsymbol{v}$ is the work done by the
pressure and $\boldsymbol{\nabla}\cdot\boldsymbol{v}$ is related to
the change in the volume, $\boldsymbol{J}_q$ contains terms related
to the heat generated by gradients of the temperature
$\boldsymbol{\nabla}T$ and dissipative forces due to viscous forces
$\sim\boldsymbol{\nabla}(\boldsymbol{\nabla}\cdot\boldsymbol{v})$
and finally but most important we have an extra term,
$\boldsymbol{\nabla}\cdot\boldsymbol{J}_\rho$, due to gradients of
the quantum potential (\ref{eq:UQ}).

 Integrating this resulting expression on a close region, we obtain
  \begin{eqnarray}
  \frac{\mathrm{d}}{\mathrm{d}t}\int n u\,\mathrm{d}V+\oint (\boldsymbol{J}_q+\boldsymbol{J}_B+p\boldsymbol{v})\cdot\boldsymbol{n}\, \mathrm{d}S
   &-&\oint\,\boldsymbol{J}_\rho\cdot\boldsymbol{n}\, \mathrm{d}S\nonumber\\
   &=&-p\frac{\mathrm{d}}{\mathrm{d}t}\int \,\mathrm{d}V\nonumber.
   \label{eq:contu2}
  \end{eqnarray}
Equation \eqref{interna} is the continuity equation for the internal
energy of the system and as usual, from here we have an expression
that would describe the thermodynamics of the system in an analogous
way as does the first law of thermodynamics, in this case for the KG
equation or a BEC. This reads
  \begin{equation}
   \mathrm{d}U=\text{\^d} Q+\text{\^d}A_Q+\text{\^d} Q_B-p\mathrm{d}V
   \label{eq:1leyBEC}
  \end{equation}
where $U=\int n u\,\mathrm{d}V$ is the internal energy of the
system, \cite{b.n}, and as we can see, its change is due to a
combination of heat $Q$ added to the system and work done on the
system. Furthermore, we have that
  \begin{equation}
   \frac{\text{\^d}A_Q}{\mathrm{d}t}=\frac{\hbar^2}{4m^2}\oint\,n(\boldsymbol{\nabla}\ln n)\dot{}\cdot\boldsymbol{n}\,
   \ \mathrm{d}S=\oint n\boldsymbol{v}_\rho\cdot\boldsymbol{n}\,\mathrm{d}S,\nonumber
   \label{eq:AQ}
  \end{equation}
is the corresponding quantum heat flux due to the quantum nature of
the KG equation. The second term on the right hand side of equation
\eqref{eq:1leyBEC} would make the crucial difference between a
classical and a quantum first law of thermodynamics.

Analogously, for the magnetic heat we have
\begin{eqnarray}
   \frac{\text{\^d}Q_B}{\mathrm{d}t}&=&\int \boldsymbol{\nabla}\cdot\boldsymbol{J}_B\,dV= \int\boldsymbol{v}\cdot(n\boldsymbol{j}_B)\,dV\nonumber\\
  &=&\frac{m}{e}\int  n\left(\frac{\partial \boldsymbol{A}}{\partial t}+(\boldsymbol{v}\cdot\boldsymbol{\nabla}) \boldsymbol{A}+(\boldsymbol{A}\cdot\boldsymbol{\nabla})\boldsymbol{v}\right)\cdot\boldsymbol{v}\,dV\nonumber\\
   \label{eq:dQ_B}
  \end{eqnarray}
where the vector potential $\boldsymbol{A}$ fulfills the Maxwell
equations, in terms of the fluxes \eqref{eq:fluxes} it reads
\begin{equation}
  {F^{\mu\nu}}_{,\nu}=-j^\mu
   \label{eq:Maxwell}
  \end{equation}
where as usual $F_{\mu\nu}=\partial_\mu A_\nu-\partial_\nu A_\mu$.
In terms of the vector and the electric potential, the Maxwell
equations are given by
\begin{subequations}
\begin{eqnarray}
  {\Box}\boldsymbol{A}&=&-\boldsymbol{j}\label{eq:Maxwellvec}\\
   {\Box}\varphi&=&-j-\frac{2en}{\kappa^2}\hat m
   \label{eq:Maxwellesc}
  \end{eqnarray}\label{eq:Maxwell2}
  \end{subequations}
where we have used the Lorentz gauge. Observe that the fluxes
contain the information of the velocity of the fluid and of the
electromagnetic term as well. This point will be important for the
superconductivity.

\section{The condensation temperature}\label{sec:TcBEC}

First, we analyze the condensation temperature in the ideal case. In this situation the associated single--particle dispersion relation is given by 
\begin{equation}
\label{E2}
E^{2}=(pc)^{2}+(mc^{2})^2.
\end{equation}
In order to compare our case with well-know critical condensation temperatures, we start analyzing the ultra--relativistic and the non relativistic cases. For these we can express the dispersion relation \eqref{E2} as $E\sim p^{s}$ \cite{Phatria}, where $s=1$ corresponds to the ultra--relativistic system and $s=2$ stands for the non--relativistic case (these topics have been extensively studied, see for example \cite{J.,Hab,Hab1,Gre,Si,Si1} and references therein). 
Taken into account the number of antiparticles, in the ultra--relativistic limit the condensation temperature is given by
\begin{equation}
\label{T1}
k_B T_{c}=\Bigl(\frac{3 \hbar ^{3} c\,n}{m}\Bigr)^{1/2},
\end{equation}
where $n$ is the charge density $n=\frac{N-\bar{N}}{V}$, $N$ is the number of particles, $\bar{N}$ corresponds to the number of anti--particles and  $V$ is the volume of the system. The quantity $N-\bar{N}$ can be written as
\begin{equation}
N-\bar{N}=\sum_{\bold{p}}[n_{\bold{p}}(\mu,T)-\bar{n}_{\bold{p}}(\mu,T) ],
\end{equation} 
where $n$ obeys the Bose--Einstein distribution function,
\begin{equation}
n_{\bold{p}}=\frac{1}{e^{\beta(E-\mu)}-1}.
\end{equation}
Here $\mu$ is the chemical potential and
$\beta=1/k_{B} T$.
Similary for the anti--bosons $\bar{n}$ we have
\begin{equation}
\bar{n}_{\bold{p}}=\frac{1}{e^{\beta(E+\mu)}-1}.
\end{equation}
The chemical potential is bounded as $|\mu|\leq mc^{2}$ and at the condensation temperature we have that $|\mu|=mc^{2}$, which corresponds to the minimum of the associated energy.
For ultra--relativistic systems without antiparticles \cite{Greiner}, the corresponding condensation temperature is given by
 \begin{equation}
\label{T2}
k_B T_{c}=\Bigl(\frac{\pi^{2}N}{ V \zeta(3)}\Bigr)^{1/3} \hbar c.
\end{equation}
Finally, in the non--relativistic case \cite{Phatria,Greiner}
\begin{equation}
\label{T3}
k_B T_{c}=\frac{2 \pi \hbar^{2}}{ m }\Bigl(\frac{N}{V\zeta(3/2)}\Bigr)^{2/3},
\end{equation}
where $\zeta(x)$ denotes the Riemann zeta function. Expressions (\ref{T1}), (\ref{T2}) and (\ref{T3}) corresponds to 3--dimensional systems, but these expressions can be generalized to different dimensions.

On the other hand, the analysis of ideal and weakly interacting non--relativistic Bose--Einstein condensates with a finite number of particles, trapped in different potentials 
(see \cite{Dalfovoro,bagnato,grossmann,Giorgini,ketterle,Haugerud,Li,zijun,Salasnich,Zobay,yukalov,yukalov1,Jaouadi,yukalov2,Phatria,Greiner,b.n,Pethick}
and references therein) shows that the main properties associated
with the condensate, in particular the condensation temperature,
strongly depends on the characteristics of the trapping potential, 
the number of spatial dimensions and the associated
single--particle energy spectrum.

In what follows we obtain the condensation temperature
associated with  our system within the semiclassical
approximation \cite{Dalfovoro,b.n,Pethick}. Inserting plane waves in the KG
equation (\ref{eq:KGT}), neglecting the term proportional to $T
^{4}$ in (\ref{eq:VT}), assuming that the temperature is
sufficiently small and considering the low velocities limit, 
allows us to obtain the  single--particle
dispersion relation between energy and momentum \cite{ZWIE}

\begin{eqnarray}
\label{SCE}
 E_{p}&\simeq &\frac{p^{2}}{2m}+\frac{\lambda}{2mc^{2}}\Big|\Phi
\Big|^{2}+\frac{\lambda}{4mc^{2}}(k_B T)^{2}\label{eq:espectro}\\ \nonumber
&+&mc^{2}
\phi+\frac{e^{2}A^{2}}{2 m c^{2}}.
\end{eqnarray}
We remain that in this work we interpret $\kappa^{2}\Big|\Phi(\vec{r},t)
\Big|^{2}$ as the spatial density $n(\vec{r},t)$ of the cloud, being
$\kappa$ the \emph{scale} of the system. Notice that if we set $\lambda=8\pi\hbar^{2}c^{2}a\kappa^{2}$ and
$\phi=\alpha r^{2}$ in \eqref{eq:espectro}, where $\alpha=1/2(\omega_{0}/c)^{2}$ and $\omega_{0}$ is a
frequency, we obtain the
semiclassical energy spectrum in the Hartree--Fock approximation for
a bosonic gas trapped in an isotropic
harmonic oscillator
\cite{Dalfovoro,b.n,Pethick}, but with two extra terms due to the
contributions of the thermal bath and to the electromagnetic field.

In this work we consider only two cases, the first one is for $A^{2}$ proportional to a constant. As we will see, in this case the associated constant can be absorbed by the chemical potential $\mu$. For the second case we consider a dependence of the form $A\sim r^{2}$.

Assuming static thermal equilibrium
$n(\vec{r},t)\approx n(\vec{r})$ \cite{Dalfovoro}, thus
\begin{equation}
\label{FE} \Big|\Phi \Big|^{2}\equiv\kappa^{-2}n(\vec{r}).
\end{equation}
The spatial density within the semiclassical approximation reads
\cite{Pethick,Dalfovoro}
\begin{equation}
\label{DE} n(\vec{r})=\frac{1}{(2 \pi \hbar)^{3}}\int  d^{3} \vec{p}
\hspace{0.1cm} n(\vec{r},\vec{p}),
\end{equation}
where $n(\vec{r},\vec{p})$ is the Bose--Einstein distribution
function given by \cite{Dalfovoro,Pethick}
\begin{equation}
\label{Sd}
 n(\vec{r},\vec{p})=\frac{1}{e^{\beta(E_{p}-\mu)}-1}.
\end{equation}
The number of particles in the 3--dimensional space obeys the
normalization condition \cite{Dalfovoro,Pethick}
\begin{equation}
\label{NC}
 N=\int d^{3}\vec{r}\hspace{0.1cm}n(\vec{r}).
\end{equation}
Integrating (\ref{DE}) over the momentum space allows to obtain the
spatial density associated to this system
\begin{equation}
\label{DE1}
n(\vec{r})=\Bigg(\frac{mk_{B}T}{2\pi\hbar^{2}}\Bigg)^{3/2}g_{3/2}(Z)
\end{equation}
where $Z=\exp [\beta(\mu-\frac{\lambda
\kappa^{-2}}{2mc^{2}}n(\vec{r})-\frac{\lambda(k_{B}T)^{2}}{4mc^{2}}-mc^{2}\phi-\frac{e^{2}A^{2}}{2mc^{2}})]$.
The function $g_{\nu}(z)$ is the so--called Bose--Einstein function
defined by \cite{Phatria}
\begin{equation}
\label{BEF}
g_{\nu}(z)=\frac{1}{\Gamma(\nu)}\int_{0}^{\infty}\frac{x^{\nu- 1}
dx}{z^{-1} e^{x}-1}.
\end{equation}
being $\Gamma(\nu)$ the Gamma function. 
In order to calculate the condensation temperature let us suppose
that our gas is trapped in a harmonic oscillator type-potential
$\phi \sim r^{2}$. Clearly, this can be extended to a more general
potentials.
Expanding (\ref{DE1}) at first order in the coupling constant
$\lambda$, using the properties of the Bose--Einstein functions
\cite{Phatria}, allows us to express the spatial density as follows
\begin{eqnarray}
\label{DE2} n(\vec{r})\approx n_{0}(\vec{r})&-&\lambda g_{3/2}(z(\vec{r}))\Bigg[\frac{\Lambda^{-6} \kappa^{-2}}{2 m c^{2}\kappa_{B}T}g_{1/2}(z(\vec{r}))\,\,\,\,\, \\\nonumber&+&\Lambda^{-3}\frac{\kappa_{B}T}{4mc^{2}}\frac{g_{1/2}(z(\vec{r}))}{g_{3/2}(z(\vec{r}))} \Bigg],
\end{eqnarray}
where
\begin{equation}
\label{ideal}
n_{0}(\vec{r})=\Lambda^{-3}g_{3/2}(z(\vec{r})),
\end{equation}
is the density for the case $\lambda=0$, $\Lambda=(2\pi \hbar^{2}/m\kappa_{B}T)^{1/2}$ is the thermal de Broglie wavelength, and $z(\vec{r})=\exp(\beta(\mu-\alpha mc^{2} r^{2}-e^{2}A^{2}/2mc^{2}))$.
In the case $A^{2} \sim r^{2}$, with the help of the normalization
condition (\ref{NC}) we obtain 
\begin{eqnarray}
\label{NPR2} N& \simeq&
\Bigl(\frac{m}{2\Omega\hbar^{2}}\Bigr)^{3/2}(k_BT)^{3}g_{3}(e^{\beta\mu})\\\nonumber&-&\frac{\lambda\kappa^{-2}m^{2}(k_BT)^{7/2}}{16\pi^{3/2}c^{2}\hbar^{6} \Omega^{3/2}}G_{3/2}(e^{\beta
\mu})\\\nonumber&-&\frac{\lambda}{4c^{2}}
\Bigl(\frac{m^{1/3}}{2\Omega\hbar^{2}}\Bigr)^{3/2}(k_BT)^{4
}g_{2}(e^{\beta \mu}),
\end{eqnarray}
where
\begin{equation}
 G_{3/2}(e^{\beta \mu})=\sum_{i,j=1}^{\infty}\frac{e^{(i+j)\beta \mu}}{i^{1/2}j^{3/2}(i+j)^{3/2}},
\end{equation}
and $\Omega=\alpha m c^{2}+const\,e^{2}/2mc^{2}$.
From expression (\ref{NPR2}), we easily obtain the case $A^{2}=
const$,
\begin{eqnarray}
\label{NPCO} N& \simeq& \Bigl(\frac{1}{2 \alpha
c^{2}\hbar^{2}}\Bigr)^{3/2}(k_BT)^{3}g_{3}(e^{\beta(\mu-\gamma)})\\\nonumber&-&\frac{\lambda\kappa^{-2}}{2c^{5}}\Bigl(\frac{m^{1/6}}{2 \pi^{1/2}\hbar^{2}\alpha^{1/2}}\Bigr)^{3}(k_BT)^{7/2}G_{3/2}(e^{\beta(
\mu-\gamma)})\\\nonumber&-&\frac{\lambda}{4mc^{5}} \Bigl(\frac{1}{2
\alpha  \hbar^{2}}\Bigr)^{3/2}(k_BT)^{4
}g_{2}(e^{\beta(\mu-\gamma)}),
\end{eqnarray}
where $\gamma$ is defined as $\gamma=const\,e^{2}/2mc^{2}$. We notice
immediately from the expressions (\ref{NPR2}) and (\ref{NPCO}) that
if the function $A^{2}$ is position dependent, the correction
in the number of particles can be associated to an effective
external potential. On the other hand, when the function $A^{2}$ is
constant, the correction can be associated to an effective
chemical potential. If we set $\lambda=0$ in  expressions (\ref{NPR2}) and (\ref{NPCO}),
we recover the expressions for the number of
particles in the non--interacting case. In the thermodynamic limit, 
in the non--interacting case
$\lambda=0$, at the condensation
temperature the value of the chemical potential is $\mu=0$
\cite{Pethick}. If we further assume that above the condensation temperature
the number of particles in the
ground state is negligible, this
allows us to obtain an expression for the condensation temperature
 $T_{0}$. For the case $A^{2}\sim r^{2}$, we obtain
\begin{equation}
\label{CTI}
k_BT_{0}=\Bigl(\frac{N}{\zeta(3)}\Bigr)^{1/3}\Bigl(\frac{2 \Omega \hbar^{2}
}{m}\Bigr)^{1/2},
\end{equation}
Analogously, for the case $A^{2}=0$ 
\begin{equation}
\label{CTI1}k_B T_{0}=\Bigl(\frac{N\sqrt{8\alpha^{3}}}{\zeta
(3)}\Bigr)^{1/3}\hbar c.
\end{equation}
Otherwise, using the properties of the Bose--Einstein functions, particularly for $g_{3}(e^{-\gamma/k_B\tilde{T}_{0}})$, when $-\gamma/k_B\tilde{T}_{0}\rightarrow 0$ \cite{Phatria}, where $\tilde{T}_{0}$ is the condensation temperature associated to the case $A^{2}=const\,\neq 0$, we obtain the shift in the condensation temperature respect to (\ref{CTI1}) caused by $\gamma \neq 0$, in function of the number of particles,
%
%\begin{equation}
%\label{CTI2} k_B
%\tilde{T}_{0}=\Bigg(\frac{N\sqrt{8\alpha^{3}}}{g_{3}( e^{- \gamma /
%k_B \tilde{T_{0}}})}\Bigg)^{1/3}\hbar c.
%\end{equation}
%
\begin{equation}
\label{SHIFT1} \frac{\tilde{T_{0}}- T_{0}}{T_{0}}\approx  \gamma
\frac{\zeta(2)}{3\zeta(3)^{2/3}\hbar c \sqrt{8\alpha^{3}}} N^{-1/3}.
\end{equation}
Clearly, the shift (\ref{SHIFT1}) vanishes when the number of particles $N \rightarrow \infty$ and
tends to the value (\ref{CTI1}). In order to obtain the leading
correction in the shift for the condensation temperature caused by the coupling constant $\lambda$ and the thermal
bath in our system, let us expand the expressions (\ref{NPR2}) and
(\ref{NPCO}) at first order in $T=T_{0}$, $\mu=0$, $\lambda=0$, and
$\gamma=0$. Additionally, at the condensation
temperature, the chemical potential in the semiclassical
approximation is given by
$\mu_{c}=\frac{\lambda\kappa^{-2}}{2mc^{2}}n(\vec{r}=0)$, as it is
suggested from expression (\ref{DE1}), thus
\begin{eqnarray}
\label{PQ1} \mu_{c}&\approx&\frac{\lambda\kappa^{-2}m^{1/2}(\kappa_{B}T_{c})^{3/2}\zeta(3/2)}{2(2\pi)^{3/2}c^{2}\hbar^{3}}\\\nonumber&-&\lambda^{3/2}\frac{\sqrt{2}\pi\kappa^{-2}(\kappa_{B}T_{c})^{2}}{(2\pi c^{2}\hbar^{2})^{3/2}}.
\end{eqnarray}
Expression (\ref{PQ1}) basically corresponds to the definition of
the chemical potential at the condensation temperature in the usual
case \cite{Dalfovoro,Pethick}, except for the extra term
contribution due to the thermal bath, and comes from the fact that
$g_{3/2}(e^{-\delta})\approx\zeta(3/2)-|\Gamma(-1/2)| \delta^{1/2}$,
when $\delta\rightarrow 0$ \cite{Phatria}. Using these facts, we finally obtain the shift in
the condensation temperature caused by $\lambda$ and the thermal
bath, in function of the number of particles in the case $A^{2}\sim
r^{2}$
\begin{eqnarray}
\label{CTR2}
\frac{T_{c}-T_{0}}{T_{0}} \equiv \frac{\Delta T_{c}}{T_{0}}=-\lambda\frac{m^{1/2}}{\kappa^{2}\hbar^{3}c^{2}}\Theta_{1}\Xi N^{1/6}
%\nonumber\\
+\lambda\Theta_{2}\Xi^{2}\,N^{1/3},\,\,\,\,\,
\end{eqnarray}
where 
\[
\Theta_{1}=\frac{1}{3\zeta(3)}\left(\frac{\zeta(3/2)\zeta(2)}{2(2\pi)^{3/2}}-G_{3/2}(1)\right),
\] 

\[\Theta_{2}=\frac{1}{3\zeta(3)}\left(\frac{1}{4\,mc^{2}}+\frac{(2\lambda)^{1/2}\zeta(2)\pi}{(2\pi)^{3/2}\kappa^{2}\hbar^{3}c^{3}}\right)
\]
and $\Xi=\left({2 \Omega\hbar^{2}}/{m}\right)^{1/4}$ with $T_{0}$ defined in (\ref{CTI}).
A similar analysis, leads us to the
shift in the condensation temperature caused by the coupling
constant and the thermal bath associated with the case $A^{2}=
const$
\begin{eqnarray}
\label{CTC}
\frac{T_{c}-T_{0}}{T_{0}} \equiv \frac{\Delta T_{c}}{T_{0}}&=&-\lambda\frac{m^{1/2}}{\kappa^{2}\hbar^{3}c^{2}}{\Theta}_{1}\tilde\Xi N^{1/6}
+\lambda{\Theta}_{2}\tilde\Xi^2N^{1/3}\,\,\,\,\,\,\,\,\,\,\, \\ \nonumber
&+&\gamma
\frac{\zeta(2)}{3\zeta(3)^{2/3}\hbar c \sqrt{8\alpha^{3}}} N^{-1/3},
\end{eqnarray}
with $\tilde\Xi=\left(2 \alpha c^{2}\hbar^{2}\right)^{1/4}$ and $T_{0}$ defined in (\ref{CTI1}).
In the case $\lambda=0$
we recover from (\ref{CTC}) the shift (\ref{SHIFT1}), as expected. From
(\ref{CTR2}) and (\ref{CTC}) we observe that the condensation
temperature $T_c$ is corrected with respect to the usual case $T_0$ 
as a consequence of the thermal bath  and 
the  contribution of the field $A^{2}$, included in the semiclassical energy
spectrum (\ref{SCE}). Additionally, we notice that the critical temperature 
associated to the symmetry
breaking (\ref{eq:Tc}) becomes very large when $\lambda\rightarrow0$
and the condensation temperatures (\ref{CTR2}) and (\ref{CTC}) tend
to the non--interacting values (\ref{CTI}) and (\ref{SHIFT1})
respectively. Setting $\alpha=1/2(\omega_{0}/c)^{2}$ and $\lambda=8\pi \hbar^{2}c^{2}\kappa^{2}a$ into expressions (\ref{CTR2}) and (\ref{CTC}) we recover the  condensation temperature for a bosonic gas trapped in an isotropic harmonic oscillator potential, but corrected by the contributions of the thermal bath and the external field $A^{2}$. For the sake of simplicity, let us analize the correction over the usual result caused by the thermal bath and the external field $A^{2}=const$ in (\ref{CTC}). For instance, in the case of  $^{87}Rb$, with $a \sim 10^{-9} m$, $N\sim10^{6}$, and $\omega \sim 10 Hz$, we obtain from the second right hand term in (\ref{CTC}) a correction up to $7.9 \times 10^{-78} \kappa^{2} + 7.5\times 10^{-38}\kappa$, and for the third right hand term, which is independient of the \emph{scale} $\kappa$, up to $const \times 10^{2}$. In other words, the \emph{scale} $\kappa$ must be very large (up to $10^{38}$) and the external field $A^{2}$ must be very weak (smaller than, or of the order of $10^{-12}$), at least near to the center of the system, in order to obtain relevant corrections over the usual result under typical conditions. For these values the symmetry breaking temperature is $\sim10^{6}K$. Finally, with the experimental data given above, we obtain for the first right hand side term of expression (\ref{CTC}) the usual shift $\sim 10^{-2}$, as expected \cite{Dalfovoro}.   
The temperatures  $T^{SB}_{c}$ and $T_{c}$ (or more
specifically, $\Delta T_{c}/T_{0}$) are related  through the
coupling constant $\lambda$, this fact could be used as a criterium
to compare both temperatures and in principle, to infer bounds
related to the \emph{scale} $\kappa$.
\section{The Phase Transition}\label{sec:PT}

From hereafter we study the transition between the $\Phi=0$ state to
the minimum $R_{min}=k_B/2\sqrt{(T^{SB}_c)^2-T^2}$ 
with $T<T^{SB}_c$.

During the time when $T>>T^{SB}_c$ there are not scalar particles in the
symmetry broken state.  Below the critical temperature
$T<T^{SB}_c$, close to the local minimum the density oscillates around
the value $n=k_B^2 \kappa^2((T^{SB}_c)^2-T^2)/4$. We study the case when
the function $S$ in (\ref{eq:psi}) has the simple expression,
$S=s_0t$, with $s_0<<mc/\hbar$ in the non-relativistic limit. This
implies that the velocity $\boldsymbol{v}=0$ as well. If there does
not exist an external force in the system, then
$\boldsymbol{F}_\phi=0$. In this case the viscosity (dissipative
term) of the BEC might in fact contain the whole information of the
phase transition. From equation \eqref{eq:sigma2} we observe that
the viscosity $\nabla\sigma$ contains only a therm with the
anisotropies of the temperature. That means that when the
temperature of the system isotropies the fluid becomes a superfluid.
Furthermore, from the expression for flux \eqref{eq:fluxesjvec} we
observe that the vectorial flux contains only a term with the vector
potential $\boldsymbol{A}$. Thus, the flux expression
\eqref{eq:fluxesjvec} becomes the London equation,
\[
\boldsymbol{j}=\frac{2ne^2}{\kappa^2}\boldsymbol{A}
\]
indicating that the system becomes superconductor. The Maxwell
equation \eqref{eq:Maxwellvec} becomes
\[
{\Box}\boldsymbol{A}=\frac{2ne^2}{\kappa^2}\boldsymbol{A}
\]
which is the Proca equation, indicating that the photon acquires a
mass $2ne^2/\kappa^2 c$. Obviously, playing with the conditions
of the system we can find situations with superconductivity or
superfluidity in different situations.

Finally, to illustrate the previous exposition, we give the
following example. Suppose that in the system there are only
condensed and excited particles of the same specie. We interpret 
$n=\kappa^2k_B^2 ((T^{SB}_c)^2-T^2)/4$ as the particles density when the symmetry $U(1)$ has been broken. At $T=0$ we expect that all particles have passed from the symmetry state into the broken symmetry one, such that the total particle density in the system is $n_T=\kappa^2k_B^2 (T^{SB}_c)^2/4$. Thus, in any time the number of particles $N_0$ in symmetry broken state is given by
  \begin{equation}
  N_0=N\, \left[1-\left(\frac{T}{T^{SB}_c}\right)^2\right],
  \label{eq:N0}
  \end{equation}
where $n/n_T=N_0/N$. Note that in this case the exponent $2$ in the critical
temperature appears naturally. As usual this expression shows the
dependence of the condensate fraction $N_0/N$ as a smooth function
of temperature from $T\sim T^{SB}_c$ down into $T=0$. In this case, the
finite temperature terms are obtained from the one loop corrections
of the SF density and are in complete agreement with the
standard theory, \cite{Hab,Si,Si1}.

Observe that the total particle density $n_T=\kappa^2k_B^2 (T^{SB}_c)^2/4$ of the scalar
particles reaching the symmetry broken state at $T=0$ can only be
determined experimentally and fix the value of the scale $\kappa$.
So in principle we are able to mimic the result that in the presence
of interactions we have $N_0<N$. The main idea we want
to point out here is that these phenomena might be equivalent for a
BEC on earth as well as for cosmological scales, so this model might be tested in the
laboratory. If confirmed, the phase transition
of a BEC can be explained using quantum field theory in a
straightforward way.

\section{Conclusions}\label{sec:conclutions}

In this work we studied the phase transition of a boson gas with zero spin represented by the KG equation with a Lagrangian containing a $U(1)$ symmetry, with mass $m$ and self-interaction $\lambda$, given by a mexican hat SF potential, immersed in a thermal bath at temperature $T$, close to the critical temperature of symmetry breaking, up to one loop in perturbations theory. We rewrite the KG equation and interpret it as a generalized Gross-Pitaevskii one at finite temperature. We show that the transition from the phase with the $U(1)$ symmetry to the phase with this symmetry broken can be interpreted as a phase transition from the gas state to the condensation state of the Bose gas. We obtained the condensation temperature as well. By rewriting the generalized Gross-Pitaevskii equation in terms of hydrodynamic quantities we were able to derive the thermodynamic of the phase transition, with viscosity and dissipative terms and find that the first law of the thermodynamics contains a new term that is a direct consequence of the quantum character of KG equation. The main result of the present work is that the phase transition do not take place at the same temperature and conditions of the condensation. We saw that, for example, Bose-Einstein condensation take place in a $^{87}$Rb crystal in normal density conditions, but that the phase transition can take place only in very hight density conditions, for example, at densities like in a neutron star. On the other hand, other materials have phase transitions and/or condensations in different conditions of density and temperature, depending on their mass, self-interaction parameter $\lambda$ and the value of the scale $\kappa$. We gave different examples.

It remains to see whether this generalization of the Gross-Pitaevskii equation can describe the transition of a Bose gas into a BEC state in the laboratory. In other words, we propose
that the superfluid and/or superconductor like-behavior in a BEC can
be measured experimentally in a laboratory, in order to compare the
results given here with realistic systems.

\acknowledgments This work was partially supported by CONACyT
M\'exico under grants 166212, 132400 and I0101/131/07 C-234/07 of the
Instituto Avanzado de Cosmologia (IAC) collaboration
(http://www.iac.edu.mx/).

\end{document}